\documentclass[prb,twocolumn,amsmath,amssymb,superscriptaddress,floatfix,nofootinbib,aps,doi=false,url=false,eprint=false,shortbibliography]{revtex4-2}
\usepackage{amsmath}
\usepackage{amsfonts}
\usepackage{graphicx}
\usepackage{dcolumn}
\usepackage{bm}
\usepackage{amsthm}
\usepackage{algorithm}
\usepackage{algpseudocode}
\usepackage[cmyk]{xcolor}
\usepackage[colorlinks,bookmarks=false,citecolor=magenta,linkcolor=magenta,urlcolor=magenta]{hyperref}
\usepackage{braket}
\usepackage{orcidlink}
\usepackage{tabularx}
\usepackage{booktabs}
\usepackage{minted}
\usepackage{quantikz}

\begin{document}
\preprint{APS/123-QED}
\title{Preparation of Fractional Quantum Hall States on Quantum Computers}

\author{Hao Wu}
\thanks{These authors contributed equally to this work.}
\affiliation{School of Physics, Beihang University, Beijing 100191, China}

\author{Lei-Yi-Nan Liu \orcidlink{0009-0005-3072-3650}}
\thanks{These authors contributed equally to this work.}
\affiliation{School of Physics, Beihang University, Beijing 100191, China}

\author{Zhao-Xin Pei}
\affiliation{School of Physics, Beihang University, Beijing 100191, China}

\author{Yi-Xuan Zhai}
\affiliation{School of Physics, Beihang University, Beijing 100191, China}

\author{Zhen-Xu Luo}
\affiliation{School of Physics, Beihang University, Beijing 100191, China}

\author{Zhao Liu}
\email{zhaol@zju.edu.cn}
\affiliation{Zhejiang Institute of Modern Physics, Zhejiang University, Hangzhou 310058, China}
\affiliation{Zhejiang Key Laboratory of Micro-Nano Quantum Chips and Quantum Control, School of Physics, Zhejiang University, Hangzhou 310027, China}

\author{Jian Cui \orcidlink{0000-0001-6643-7625}}
\email{jiancui@buaa.edu.cn}
\affiliation{School of Physics, Beihang University, Beijing 100191, China}

\date{\today}

\begin{abstract}
The realization of fractional quantum Hall (FQH) states, characterized by fractional charge and intrinsic topological order, on quantum computers represents a central challenge at the interface of condensed matter physics and quantum information science. Current methods are grouped into two types: methods based on (quasi-)adiabatic evolution of complex parent Hamiltonians to yield target states, and circuit-based approaches for direct state preparation, which are confined to effectively one-dimensional systems near the thin cylinder or torus limit.
We introduce a complementary scheme relying on direct quantum circuit construction, which works for arbitrary geometries.
Specifically, we present a method to precisely prepare the $\nu=1/3$ Laughlin state on the sphere geometry and demonstrate that it significantly reduces the required number of two-qubit gates and circuit depth, compared to variational quantum circuit approaches. In addition, we employ optimal control techniques to design control pulses for both superconducting and Rydberg atom platforms, identifying experimentally feasible protocols for state preparation. Our results provide an efficient and hardware-relevant pathway for realizing generic FQH states on both noisy intermediate-scale and fault-tolerant quantum devices. 
\end{abstract}

\maketitle

\section{Introduction}

The quantum Hall effect provides a paradigmatic example of phases of matter beyond the Landau symmetry-breaking paradigm, where global topological properties rather than local order parameters become the main focus~\cite{Klitzing2020}. While the integer quantum Hall effect admits a single-particle description in terms of topological invariants of Landau levels, the fractional quantum Hall (FQH) effect arises purely from strong interactions and realizes intrinsically correlated topological states with long-range entanglement~\cite{Laughlin1983,Nakamura2020,de-Picciotto1997,Jiang2017,Wen2008}. These states host quasiparticles with fractional charge and anyonic statistics, including non-Abelian excitations whose braiding enables fault-tolerant quantum computation~\cite{Kim2026,Das2005,Bravyi2006,Barkeshli2014,Mong2014,Nayak2008}. 
While topologically ordered states have been realized on various quantum simulation platforms, most experimentally accessible examples rely on exactly solvable models such as the toric code and surface code~\cite{Satzinger2021,Bluvstein2022,Semeghini2021}. In contrast, FQH states emerge from strongly interacting systems without simple implementable parent Hamiltonians, making their investigation on quantum computers a fundamentally more challenging and largely open problem.

Recent years have witnessed significant progress toward realizing FQH states in both analog and digital quantum devices. Among these, the Laughlin states, the first theoretically proposed and verified explanation of the FQH effect, serve as the simplest and most fundamental representative~\cite{Laughlin1983}. 
Existing methods fall into two categories: those that generate target states via (quasi-)adiabatic evolution of parent Hamiltonians, and those that construct quantum circuits to directly prepare the desired quantum states. 
In the former, several recent approaches construct the target Laughlin states as the ground states of effective models using Hamiltonian-based variational ans\"atze or adiabatic evolution. These have been experimentally implemented recently demonstrating the simulation of dynamical properties, collective excitations, and anyon braiding of FQH states~\cite{Kirmani2022,Kirmani_2023}. In parallel, advances in programmable platforms such as ultracold atoms, photonic systems, and superconducting circuits have enabled the realization of few-body FQH states and the observation of their characteristic signatures~\cite{Wang2024,Leonard2023}. 
In the latter, efficient quantum algorithms have been proposed to generate Laughlin states for systems in the one dimentional limit using linear-depth circuits~\cite{Rahmani2020}, and a variety of state-preparation strategies, including variational quantum circuits, optimal control techniques, and machine-learning-assisted protocols, have been developed to improve the fidelity and efficiency of preparing FQH states~\cite{Lingnan2025,Blatz2024,Wu2025}. Though physically well-motivated, the former approach requires the implementation of multipartite interactions, notoriously hard to achieve in experiments, and relies on very deep quantum circuits (or long evolution times) for adiabaticity, both of which present major experimental challenges. Moreover, such Hamiltonian-based constructions are typically tailored to specific models and geometries, which may limit their flexibility and applicability on general-purpose quantum devices. 
The latter are either constrained to effectively one-dimensional systems near the thin cylinder or torus limit, or involve a large number of multi-qubit gates and deep circuits. 
Thus, achieving efficient, scalable, and hardware-friendly preparation of FQH states on quantum computers therefore remains an outstanding challenge.

Here, we explore the preparation of FQH states on quantum computers in a geometry that is isotropic and free from edge effects. 
Specifically, we take the $\nu=1/3$ FQH Laughlin state on the sphere with seven orbitals in the lowest Landau level as a concrete example, and construct a quantum circuit that directly prepares this state, without relying on the implementation or approximation of a parent Hamiltonian. Our approach exploits the sparsity of the FQH wavefunction, significantly reducing the required gate resources. Moreover, the method is general and can be extended to other geometries and larger systems. 
This differs from most existing theoretical proposals, which are typically formulated near the thin-cylinder (or thin-torus) limit~\cite{Rahmani2020,Kirmani2022}, where the circumference of the cylinder is much smaller than its length. In this quasi-one-dimensional regime, the ground state of the effective FQH Hamiltonian tends to a charge-density-wave (CDW) configuration $\ket{100100\cdots}$. Close to this limit, the ground state can be systematically constructed from the CDW seed configuration using simple squeezing rules, where particles are locally rearranged while preserving the total momentum~\cite{Nakamura2012}. As a result, the wavefunction acquires a relatively simple hierarchical structure that can be efficiently described within a restricted Hilbert space. However, such quasi-one-dimensional descriptions do not fully capture the essential features of the Laughlin state in two dimensions. In particular, key signatures of topological order, including long-range entanglement and the universal structure of the entanglement spectrum, are not faithfully represented in these approximate descriptions.
In contrast, our construction directly describes the FQH state in a genuine two-dimensional geometry, where the wavefunction exhibits a more intricate and nonlocal squeezing structure. The circumference of the great circle of the sphere grows with the system size and exceeds those reached near the thin-cylinder limit, ensuring that no quasi-one-dimensional reduction applies. Furthermore, rather than Hamiltonian-based or adiabatic evolution schemes~\cite{Blatz2024,Wu2025}, we directly construct a quantum circuit targeting the desired FQH state. This approach bypasses constraints associated with gap protection and long evolution times, and provides a flexible and scalable route to preparing FQH states. Our results demonstrate that the resulting state faithfully captures the essential features of the $\nu=1/3$ FQH phase, highlighting a complementary pathway toward realizing strongly correlated topological states on quantum devices.

This paper is organized as follows. In Sec.~II, we introduce the target FQH state considered in this work, defined for three particles in seven orbitals on the spherical geometry. In Sec.~III, we present our approaches to preparing this state on quantum computers, including direct circuit construction, variational quantum circuits, and optimal-control-based methods, and benchmark their performance in terms of state-preparation fidelity. 
In Sec.~IV, we incorporate realistic experimental considerations and analyze the robustness of our protocols in the presence of noise.
In Sec.~V, we characterize the prepared states using experimentally accessible figures of merit beyond fidelity, including the entanglement spectrum, which can be obtained via subsystem tomography and serves as a direct signature of the underlying topological order, the density profiles, and the many-body correlation functions.
Finally, in Sec.~VI, we discuss the implications of our results and outline future directions. While our primary focus is on the seven-qubit case, we also explore the extension to larger systems, such as ten qubits, where the preparation becomes significantly more challenging. Addressing these challenges and achieving scalable preparation of larger FQH states constitute an important direction for future work.

\section{The $\nu=1/3$ Laughlin state}
When a strong perpendicular magnetic field is applied to a two-dimensional electron gas, the kinetic energy of electrons is quenched and the system forms a set of highly degenerate Landau levels. Within each Landau level, the kinetic energy is frozen, and the electron-electron interaction becomes the dominant energy scale that determines the many-body ground state. Among these, the $\nu=1/3$ FQH state is described by the Laughlin wavefunction~\cite{Laughlin1983}, which captures the essential features of a strongly correlated topological phase. This state exhibits long-range entanglement and supports fractionalized excitations with anyonic statistics~\cite{Wen1990,Wen2002,Wen2008,Jiang2017,Nayak2008}.

We consider the FQH state on the sphere geometry with full rotational symmetry~\cite{HaldaneSphere1983}, where there are $2Q$ magnetic flux quanta through the surface of the sphere. In the lowest Landau level, the single-particle wavefunctions take the form of 
\begin{eqnarray}
   \phi_{m}=\mathcal{N}_m
(-1)^{Q-m}u^{Q+m}v^{Q-m},
\end{eqnarray}
where $\mathcal{N}_m=\left[\frac{2Q+1}{4\pi}\frac{(2Q)!}{(Q-m)!(Q+m)!}\right]^{1/2}$, $u=\cos(\theta/2)e^{i\phi/2}$, $v=\sin(\theta/2)e^{-i\phi/2}$, and 
$\theta\in[0,\pi]$ and $\phi\in[0,2\pi)$ are the spherical coordinates. 
Each single-particle state is labeled by the angular momentum $m=-Q,-Q+1,\cdots,Q-1,Q$. The norm of the $m$th state reaches the maximum at $\cos\theta=m/Q$, thus we can imagine it as an orbital along the latitude line of the sphere. For the case of three particles in seven orbitals, the $\nu=1/3$ Laughlin state can be expressed in the occupation-number basis as 
\begin{eqnarray}
    \displaystyle \ket{\Psi_{\mathrm{FQH}}} & =\frac{1}{\sqrt{7}} \Big( \ket{1001001} + \ket{0101010} - \ket{0011100} \nonumber \\
     & - \sqrt{2}\ket{1000110}  - \sqrt{2}\ket{0110001} \Big), \label{target_state}
\end{eqnarray}
where the coefficients can be calculated via Jack polynomials~\cite{Bernevig2008}. Although the expansion in Eq.~(\ref{target_state}) is also composed exclusively of the root configuration $\ket{100100\cdots}$ and its inwardly squeezed descendants, the underlying squeezing rule is non-local. This stands in contrast to the local squeezing rule with uniform amplitudes near the thin-cylinder (and thin-torus) limit~\cite{Nakamura2012,Rahmani2020,Kirmani2022}. Such complexity makes the spherical formulation significantly more challenging, yet also more faithful to the original description of the FQH effect. In the following, we will design three approaches to prepare Eq.~(\ref{target_state}) on quantum devices.

\section{Preparation methods and results}

\subsection{Digital circuit}

\begin{figure*}
    \includegraphics[width=\linewidth]{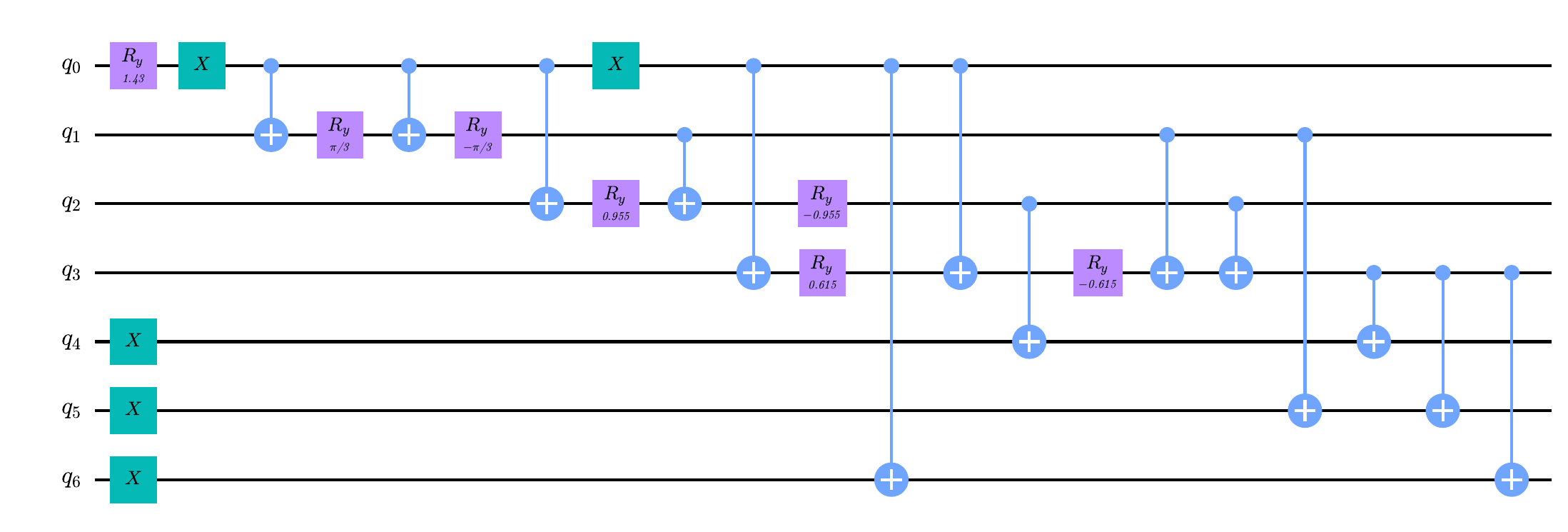}
    \caption{Circuit that realizes the target state defined in Eq.~\eqref{target_state}, depicted using qiskit~\cite{qiskit2024}. There are 12 single qubit gates and
    14 CNOT gates in this circuit. 
    The circuit is initialized in $\ket{0}^{\otimes 7}$. }
    \label{circuit}
\end{figure*}

Our first method is to directly design an exact low-depth quantum circuit that prepares the FQH state in Eq.~\eqref{target_state}. This construction is based on the binary-tree (or decision-diagram) representation of a quantum state~\cite{Long2001, Zhang2022}, in which the target wavefunction is recursively decomposed into a hierarchy of conditional amplitudes. Within this framework, the state preparation proceeds qubit by qubit: each qubit is initialized through a sequence of controlled rotations whose angles are determined by the relative weights of different branches in the tree. The binary-tree scheme to prepare an arbitary quantum state was originally proposed in Ref.~\cite{Long2001}, where the operation on the $n$-th qubit generally depends on all preceding $n-1$ qubits, leading to multi-controlled gates of the form $C^{n-1}U$, which become increasingly costly as the system size grows. More recent approaches~\cite{Zhang2022} exploit parallelization and ancillary qubits to reduce the circuit depth to its optimal scaling by performing operations on different branches simultaneously, at the expense of substantial qubit overhead.

Here, our implementation adopts a sequential construction without introducing ancillary qubits, but leverages the intrinsic structure of the FQH state to significantly simplify the control requirements. Owing to the constrained structure of the wavefunction in the occupation-number basis (only the root configuration and its inward squeezing are present), the dependence on preceding qubits is highly restricted. As a result, the state preparation can be implemented using far fewer control qubits than in the generic case. For the seven-qubit FQH state considered here, all required operations can be decomposed into $12$ single-qubit rotations and $14$ CNOT gates. For larger systems, such as the ten-qubit case, the construction requires at most three control qubits, corresponding to $C^3U$ operations. We give a minimal example using a three qubit state in Appendix~\ref{q3example}.

We present our circuit for the precise preparation of the state $\ket{\Psi_{\mathrm{FQH}}}$, up to a global $\pi$ phase in Fig.~\ref{circuit}. 
The circuit that produces more complicated ten-qubit $\nu=1/3$ FQH state is shown in the Appendix~\ref{fqh10_circuit}. 

\subsection{Variational Circuit}

We next employ a variational quantum circuit (VQC) approach and contrast it with the direct circuit construction described in Part A. Variational quantum algorithms have been widely applied across a range of quantum information tasks~\cite{Cerezo2021}, and have also been proposed as a flexible framework for preparing nontrivial quantum states on quantum devices~\cite{Bornens2025,Consiglio2025,Zuniga2024,Bond2025,Kuzmin2020}. In this framework, the target FQH state is approximated by a parameterized quantum circuit, whose parameters are optimized to maximize the fidelity with respect to the target state. This variational approach provides a flexible alternative to exact circuit constructions, particularly when an explicit decomposition of the target state becomes challenging.

The ansatz circuit is constructed based on the chosen qubit topology of the resource platform and consists of repeated layers of parameterized single-qubit rotations and entangling operations. Specifically, each composite layer contains two rotation blocks, implemented as parametrized independent single-qubit rotations on all qubits, separated by an entangling block composed of nearest-neighbor two-qubit gates. This layered structure provides a balance between expressibility and circuit depth, allowing the ansatz to capture the essential correlations of the target state while remaining compatible with realistic hardware constraints. In our simulations, we employ a variable number of composite layers to prepare the quantum state.

We denote the $n$-th layer as $U(\boldsymbol{\alpha}_n,\boldsymbol{\beta}_n)$, where $\boldsymbol{\alpha}_n$ and $\boldsymbol{\beta}_n$ are the rotation angles associated with the $R_y$ and $R_x$ gates in that layer, respectively. Starting from the initial state $\ket{\psi_0}=\ket{0}^{\otimes 7}$, the output state after $N$ layers is given by
\begin{equation}
    \ket{\psi_N} = U(\boldsymbol{\alpha}_N,\boldsymbol{\beta}_N) \cdots
    U(\boldsymbol{\alpha}_2,\boldsymbol{\beta}_2)
    U(\boldsymbol{\alpha}_1,\boldsymbol{\beta}_1)\ket{\psi_0}.
\end{equation}
The quality of the prepared state is quantified by the fidelity with respect to the target FQH state,
\begin{equation}
    \mathcal{F}_N(\boldsymbol{\alpha}_1\cdots\boldsymbol{\alpha}_N;\boldsymbol{\beta}_1\cdots \boldsymbol{\beta}_N)
    = \bigl|\braket{\Psi_{\mathrm{FQH}}|\psi_N}\bigr|^2. 
\end{equation}
The circuit parameters are optimized by maximizing this fidelity using the AdaBelief optimizer~\cite{Juntang2020}, with gradients computed via automatic differentiation implemented in JAX~\cite{jax2018}.

\begin{figure}
    \centering
    \includegraphics[width=0.9\linewidth]{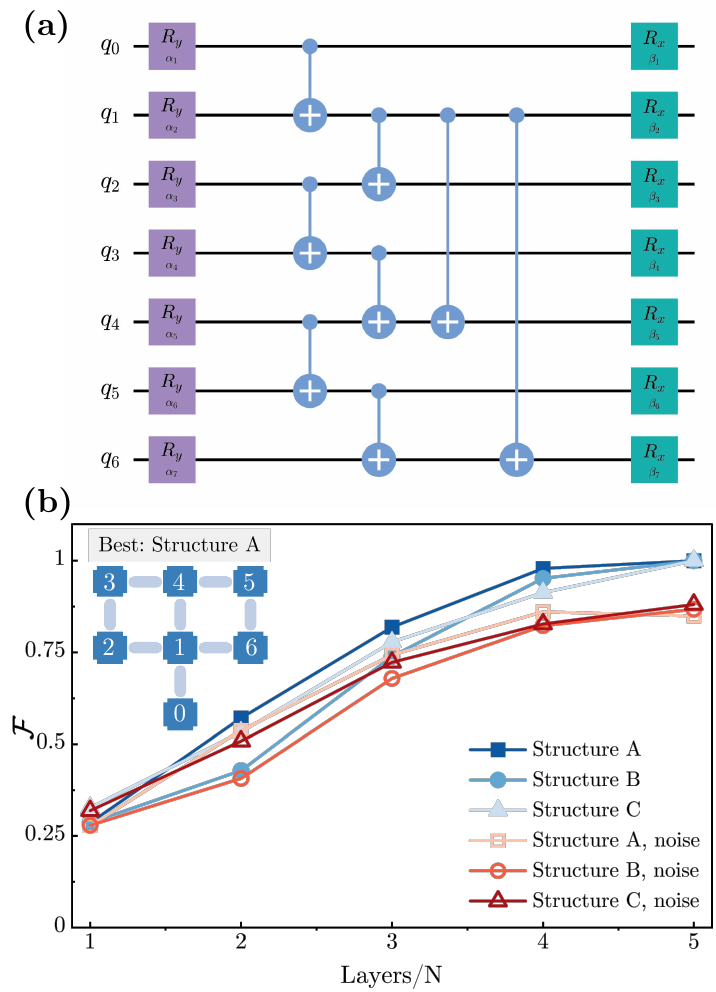}
    \caption{VQC results. (a) One-layer VQC circuit for Structure A, where the $R_x$ rotation layer in the $n$-th layer can be regarded as a single unit with the $R_y$ rotation layer in the $(n+1)$-th layer forming a structure of one rotation layer. (b) Fidelity as a function of the number of layers, with the noise definition detailed in Sec.~\ref{sec:experimental_considerations} and the noise strength set to $p_1=0.0002$ and $p_2=0.004$.  The maximum achievable fidelity increases with circuit depth. Among the considered ansatz structures, Structure A attains the highest fidelity, while results for other structures are discussed in Appendix~\ref{app:vqe_parameters}.}
    \label{fig:topology}
\end{figure}

We then incorporate hardware constraints motivated by superconducting quantum platforms,
where two-qubit gates are typically limited to nearest-neighbor connectivity. 
For the seven-qubit system considered here, we design a two-dimensional planar topology that maximizes the number of available nearest-neighbor two-qubit interactions, as shown in the inset of Fig.~\ref{fig:topology}(b). Based on this connectivity, we construct the entangling layers using all allowed nearest-neighbor two-qubit gates, resulting in up to eight two-qubit interactions per layer, arranged into a circuit of depth three, as illustrated in Fig.~\ref{fig:topology}(a). 

To evaluate the performance of the ansatz, we systematically study circuits with different numbers of composite layers. As shown by the blue solid squares in Fig.~\ref{fig:topology}(b), the circuit with $4$ composite layers achieves a fidelity of $97.9\%$, while the $5$-layer circuit further improves the fidelity to $99.99\%$, realizing an almost perfect FQH state. In terms of parameter scale, each composite layer contains $14$ independent parameters. The $4$-layer circuit reaches a high fidelity with only $56$ free parameters and $32$ CNOT gates. This reflects the nontrivial entanglement structure of the target FQH state, which requires sufficient circuit depth to be faithfully captured. For the $10$-qubit \( \nu=1/3 \) Laughlin state, a state preparation fidelity of 83.78\% is achieved via our VQC approach, with full details provided in Appendix~\ref{fqh10_circuit}.

We also investigate alternative qubit topologies, including a modified two-dimensional layout (structure B) and a one-dimensional chain(structure C), as shown in Appendix~\ref{app:vqe_parameters}. In both cases, the achieved fidelity at a fixed circuit depth is lower than that of the optimized planar topology. This indicates that the choice of qubit connectivity plays an important role in the efficiency of the variational optimization. In particular, more connected topologies enable a larger number of effective two-qubit interactions, thereby enhancing the expressibility of the ansatz. This effect is expected to become even more significant for larger system sizes.

Finally, we compare the variational approach with the direct circuit construction. Although the VQC method provides a flexible and hardware-adaptable framework, it generally requires deeper circuits and a larger number of two-qubit gates to achieve comparable fidelity. In particular, the increased number of CNOT gates leads to a higher sensitivity to noise on superconducting platforms. In contrast, the direct circuit construction exhibits a clear advantage in terms of gate count, making it more favorable for near-term implementations. The optimized parameters for the VQC circuit with four composite layers are provided in Appendix~\ref{app:vqe_parameters}.

\subsection{Optimal Control}

We next consider an optimal control approach and contrast it with previous theoretical proposals based on adiabatic state preparation. In conventional schemes, the target FQH state is obtained by slowly evolving the system under a time-dependent Hamiltonian, starting from a simple initial state, while optimal control techniques are employed to enhance the fidelity of the adiabatic process \cite{Blatz2024,Wu2025}. Such approaches inherently rely on maintaining adiabaticity and typically require long evolution times to suppress unwanted excitations.

In contrast, our approach does not rely on adiabatic evolution. Instead, we directly search for an optimal control protocol that prepares the target FQH state starting from a simple product state. Specifically, we initialize the system in a trivial state $\ket{\psi_0}$ and optimize the time-dependent control fields $f_i(t)$ in the Hamiltonian
\begin{equation}
\hat{H}(t) = \hat{H}_0 + \sum_{i=1}^n f_i(t)\hat{H}_i,
\end{equation}
such that the overlap with the desired state at the final time is maximized. The corresponding objective function is given by
\begin{equation}
\mathcal{F}[f_1(t),\cdots,f_n(t)] = |\braket{\Psi_{\mathrm{FQH}} | U(T) | \psi_0}|^2,
\end{equation}
where
\begin{equation}
U(T) = \mathcal{T}\exp\left[-i\int_{0}^{T}\hat{H}(\tau)\ \mathrm{d}\tau\right]
\end{equation}
is the time-evolution operator generated by $\hat{H}(t)$.
To solve this optimal control problem, we employ the dressed chopped random basis (dCRAB) method, which provides an efficient parametrization of the control fields and enables gradient-free optimization in high-dimensional landscapes \cite{Rach2015,Sorensen2018,Muller2022}. 

We apply this optimization framework to two leading quantum simulation platforms, superconducting system and Rydberg atom arrays. For superconducting systems, the control fields correspond to microwave drives. For Rydberg atom platforms, the controls are implemented through time-dependent laser parameters, including Rabi frequencies and detunings, which govern the effective many-body dynamics. In both cases, we optimize physically realistic control waveforms under experimentally relevant constraints, and evaluate the achievable fidelity of preparing the target FQH state.

\subsubsection{Superconducting system}
The full Hamiltonian for the superconducting system is given by $\hat{H}^{\mathrm{sc}}(t)=\hat{H}_0^{\mathrm{sc}}+\hat{H}_c^{\mathrm{sc}}(t)$, where
\begin{eqnarray}
    \hat{H}_0^{\mathrm{sc}} &=& \sum_{i=1}^{N-1} J_{i,i+1}\left(\hat{\sigma}_i^{+}\hat{\sigma}_{i+1}^{-} + \hat{\sigma}_i^{-}\hat{\sigma}_{i+1}^{+}\right), \\
    \hat{H}_c^{\mathrm{sc}}(t) &=& \sum_{i=1}^{N}\left[\omega_{i}(t)\hat{n}_i + g_i(t)\hat{\sigma}^x_i\right],
\end{eqnarray}
with $N$ the number of qubits, $\hat{n}_i=\hat{\sigma}^+_i\hat{\sigma}^-_i$ the number operator and $\hat{\sigma}^x_i=\hat{\sigma}^{+}_i+\hat{\sigma}^{-}_i$ is the Pauli operator in $x$ direction. Here $J_{i,i+1}$ denotes the nearest-neighbor coupling strength, while $\omega_i(t)$ and $g_i(t)$ represent the time-dependent detuning and driving amplitude applied to the $i$th qubit, respectively. We adopt the nearest-neighbor couplings: $J_{1,2}=27.6\ \mathrm{MHz}$, $J_{2,3}=27.4\ \mathrm{MHz}$, $J_{3,4}=29.1\ \mathrm{MHz}$, $J_{4,5}=27.6\ \mathrm{MHz}$, $J_{5,6}=26.5\ \mathrm{MHz}$, $J_{6,7}=29.2\ \mathrm{MHz}$~\cite{doi:10.1126/science.aay0600}.

The fidelity thus depends on the control fields $\omega_i(t)$ and $g_i(t)$. To make the optimization tractable, we parametrize these time-dependent control pulses using a truncated trigonometric basis, which reduces the functional optimization problem $\mathcal{F}[\omega_i(t), g_i(t)]$ to a finite-dimensional optimization over a set of parameters $\boldsymbol{x}$\cite{PhysRevA.92.062343}. The resulting multivariable objective function $\mathcal{F}(\boldsymbol{x})$ is then maximized using a gradient-based optimization method. 

\begin{figure*}
    \includegraphics[width=\linewidth]{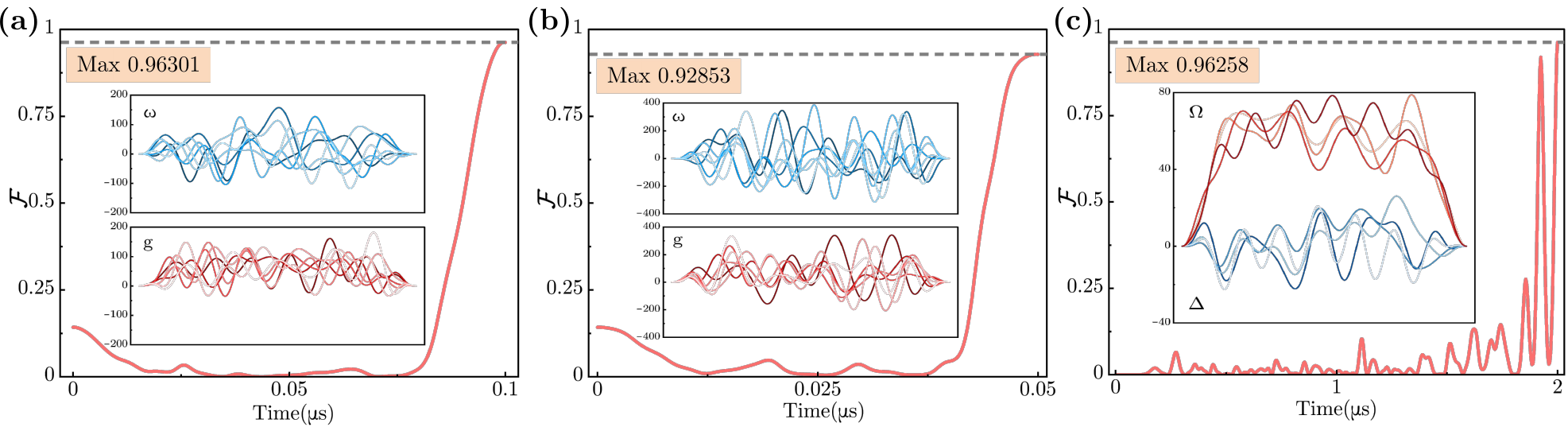}
    \caption{Fidelity of analog waveform optimization. (a,b) Fidelity as a function of optimization iteration and evolution time on a superconducting platform for pulse durations of $0.05\ \mu\mathrm{s}$ and $0.10\ \mu\mathrm{s}$, respectively. (c) Corresponding results for the Rydberg atom array. Inset: Waveform plot of the optimization results.}
    \label{fig:analog}
\end{figure*}

Using the numerical optimization framework described above, we perform state preparation for an $N=7$ qubit system. To investigate the role of the total evolution time, an important physical resource, in determining the achievable fidelity, we carry out independent optimization runs for different evolution durations. 

The numerical results demonstrate that the dCRAB optimization algorithm converges reliably across a range of evolution times. Notably, increasing the total evolution time leads to a significant improvement in the achievable fidelity. For example, when the evolution time is increased from $0.05\ \mu\mathrm{s}$ to $0.10\ \mu\mathrm{s}$, the final fidelity improves from $92.85\%$ to $96.30\%$,  as illustrated in Fig.~\ref{fig:analog}(a)(b).
These results highlight a clear trade-off between evolution time and control performance in quantum optimal control. A longer evolution time effectively enlarges the accessible control landscape, enabling smoother dynamical trajectories that more closely approach the target state while suppressing unwanted excitations.

\subsubsection{Rydberg atom arrays}

For the Rydberg atom platform, we consider a one-dimensional array of $N$ atoms. The system dynamics is governed by the full many-body Hamiltonian $\hat{H}^{\mathrm{Ryd}}(t)$, which includes the laser driving terms and the long-range van der Waals interactions, 
\begin{eqnarray}
    \hat{H}^{\mathrm{Ryd}}(t)&=&\hat{H}_0^{\mathrm{Ryd}}+\hat{H}_c^{\mathrm{Ryd}}(t)\\
    H_0^{\mathrm{Ryd}} &=& \sum_{i<j}\frac{C_6}{d^6|i-j|^6}\hat{n}_i\hat{n}_j, \\
    H_c^{\mathrm{Ryd}}(t) &=& \sum_{i=1}^{N} \left[ \frac{\Omega_i(t)}{2}\hat{\sigma}^x_i - \Delta_i(t) \hat{n}_i \right].
\end{eqnarray}
where $C_6$ is the interaction coefficient, $d$ is the spacing between adjacent atoms, and $\Omega_i(t), \Delta_i(t)$ represent the Rabi frequency and detuning for the $i$-th atom, respectively. We adopt $C_6 = 611\ \mathrm{GHz}\cdot\mu m^6$ corresponding to the $|68S_{1/2}\rangle$ Rydberg state of $^{87}\text{Rb}$ atoms, and a fixed nearest-neighbor atomic spacing $d = \ 6 \mu m$~\cite{SIBALIC2017319}. 

To reduce the complexity of the high-dimensional search space while leveraging the inversion symmetry of the target Laughlin state, we impose a symmetric constraint on the control pulses, 
\begin{equation}
    \Omega_i(t) = \Omega_{N-i+1}(t), \quad \Delta_i(t) = \Delta_{N-i+1}(t).
\end{equation}
For a system size of $N=7$, this strategy reduces the number of independent control channels from $14$ to $8$. 
The atoms are addressed in symmetric pairs (e.g., $\{1,7\}, \{2,6\}, \{3,5\}$) with the central atom $\{4\}$ controlled independently.

We applied the dCRAB algorithm to optimize these symmetric channels. The optimization results are shown in Fig.~\ref{fig:analog}(c), demonstrating that the algorithm converges efficiently and achieves a final fidelity of $96.26\%$, which is of the same order of magnitude as the optimal result of superconducting system evolution. The inset of Fig.~\ref{fig:analog}(c) presents the time dependence of the fidelity during the evolution, revealing pronounced oscillations in the final stage of the protocol.

Further numerical analysis indicates that these late-time fidelity oscillations mainly arise from the rapid evolution of the relative phases among the dominant components associated with the target state. Within this late-time interval, the probability distribution of the principal computational-basis components has already become approximately stable, whereas their relative phases continue to evolve. Since the fidelity reflects the interference between the instantaneous state and the target state, even when the populations of the dominant components remain nearly unchanged, deviations of the relative phases from the target-state phase relations suppress constructive interference and lead to a rapid decrease in fidelity. When these phases become realigned, the interference is enhanced again and the fidelity correspondingly rises. In other words, the high-frequency peaks and dips near the end of Fig.~\ref{fig:analog}(c) primarily reflect the alternating matching and mismatching of the relative phases among the dominant target-state components.

\section{Noise analysis}\label{sec:experimental_considerations}
Next, we investigate the impact of noise on the performance of the above state-preparation methods. For the exact circuit and variational quantum circuit (VQC) approaches, we adopt a gate-level independent depolarizing noise model, which serves as a standard benchmark for assessing the robustness of NISQ algorithms and is compatible with the noise calibration frameworks of current quantum hardware platforms~\cite{Nielsen2010,Preskill2018,qiskit2021,Escofet2025}. Specifically, after each ideal unitary operation, we apply a depolarizing channel, assuming statistically independent noise for each gate.
For single-qubit gates, the depolarizing channel is defined as
\begin{equation}
\mathcal{E}_{1}(\rho) = (1-p_1)\rho + \frac{p_1}{3}\sum_{i=1}^{3}\sigma_i\rho\sigma_i,
\end{equation}
where $p_{1}$ denotes the single-qubit error probability and ${\sigma_i}$ are the Pauli matrices. 
For two-qubit CNOT gates, the depolarizing channel takes the form
\begin{equation}
\mathcal{E}_{2}(\rho) = (1-p_{2})\rho + \frac{p_{2}}{15}\sum_{\substack{i,j=0\\(i,j)\neq (0,0)}}^3 (\sigma_i\otimes\sigma_j) \rho (\sigma_i\otimes\sigma_j),
\end{equation}
where $p_{2}$ denotes the two-qubit error probability.

\begin{figure}
    \centering
    \includegraphics[width=\linewidth]{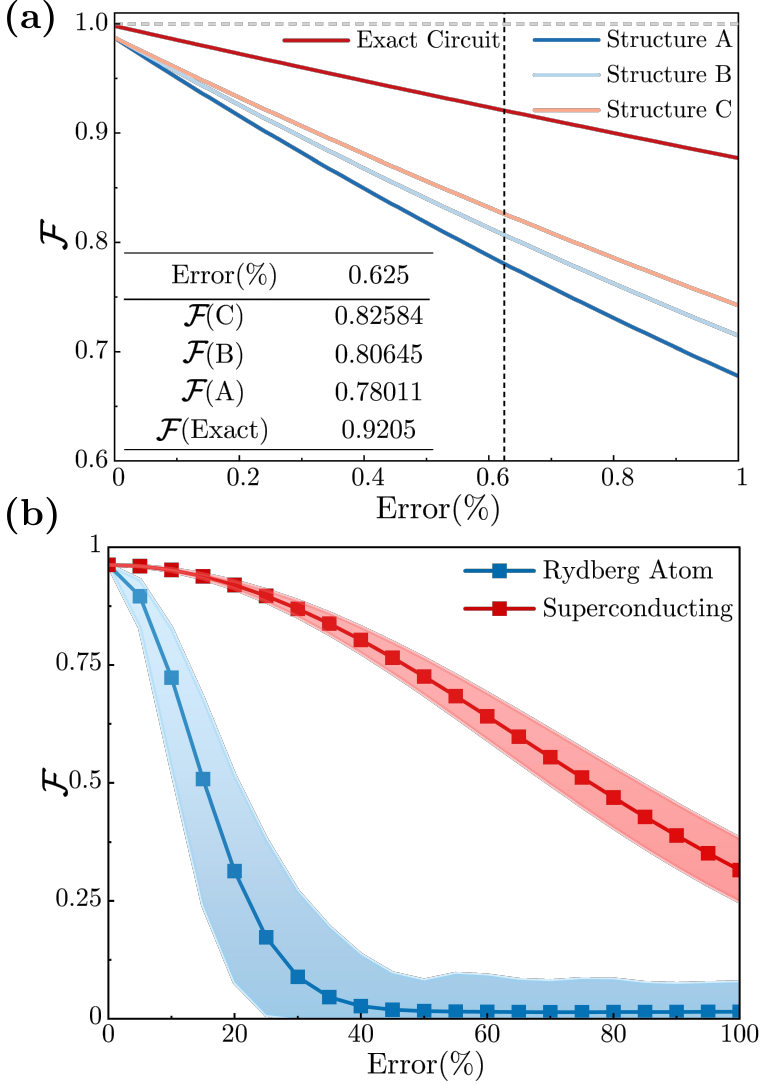}
    \caption{Fidelity under noise. (a) Fidelity of the VQC and exact circuit as a function of the two-qubit error rate $p_2$, evaluated using density-matrix simulations (no error bars). (b) Fidelity of the analog method versus noise strength $\epsilon$. Error bars indicate statistical fluctuations arising from the sampling process. \label{fig:all_noise}}
\end{figure}

State-of-the-art quantum hardware typically achieves near-perfect single-qubit gate fidelities, while two-qubit gate errors remain the dominant source of decoherence~\cite{Preskill2018,Eisert2025}. We therefore focus on the impact of two-qubit noise. As shown in Fig.~\ref{fig:all_noise}(a), we fix the single-qubit error rate at $0.02 \%$ and plot the state-preparation fidelity as a function of $p_{2}$. All schemes exhibit a monotonic decrease in fidelity with increasing two-qubit noise strength, as expected.
Notably, the exact circuit exhibits the slowest fidelity degradation, whereas the VQC (Structure A) shows the fastest decay. This behavior can be directly attributed to differences in circuit depth and two-qubit gate count. These results highlight the critical role of circuit complexity in practical implementations. For a representative state-of-the-art two-qubit gate fidelity of $99.5\%$, which corresponds to a two-qubit depolarizing error rate $p_2=0.625\%$~\cite{PhysRevLett.134.090601,Eisert2025}, the exact circuit still retains a state preparation fidelity above $92\%$. This result highlights the strong noise robustness of our direct circuit construction and its promising implementation feasibility on near-term quantum processors.

For the optimal control schemes, we model amplitude noise in the optimized control waveforms, corresponding to background fluctuations and amplitude jitter in microwave or laser control lines. Let $u(t)$ denote the ideal control waveform and $A=\max_t |u(t)|$ its maximum amplitude. The noisy waveform is modeled as 
\begin{equation}
    \tilde{u}(t) = u(t) + A\delta(t), \quad 
\delta(t) \sim \mathcal{U}\left(-\varepsilon,\varepsilon\right)
\end{equation}
where $\varepsilon$ characterizes the noise strength and $\mathcal{U}(a,b)$ denotes a uniform distribution over the interval $[a,b]$. 

As shown in Fig.~\ref{fig:all_noise}(b), the superconducting system exhibits stronger robustness against amplitude noise compared to the Rydberg atom system under the same relative noise strength. The superconducting case (red curve) shows a gradual decrease in fidelity, maintaining approximately $32\%$ fidelity even at $\varepsilon=100\%$. In contrast, the Rydberg system (blue curve) exhibits a much sharper decay, with the fidelity dropping to around $20\%$ already at $\varepsilon\approx 30\%$. This indicates that the Rydberg platform is more sensitive to amplitude fluctuations and requires tighter control precision to achieve high-fidelity state preparation.

\section{Figures of merit other than fidelity}

\subsection{Entanglement spectrum}

\begin{figure}[!h]
  \centering
  \includegraphics[width=\linewidth, keepaspectratio]{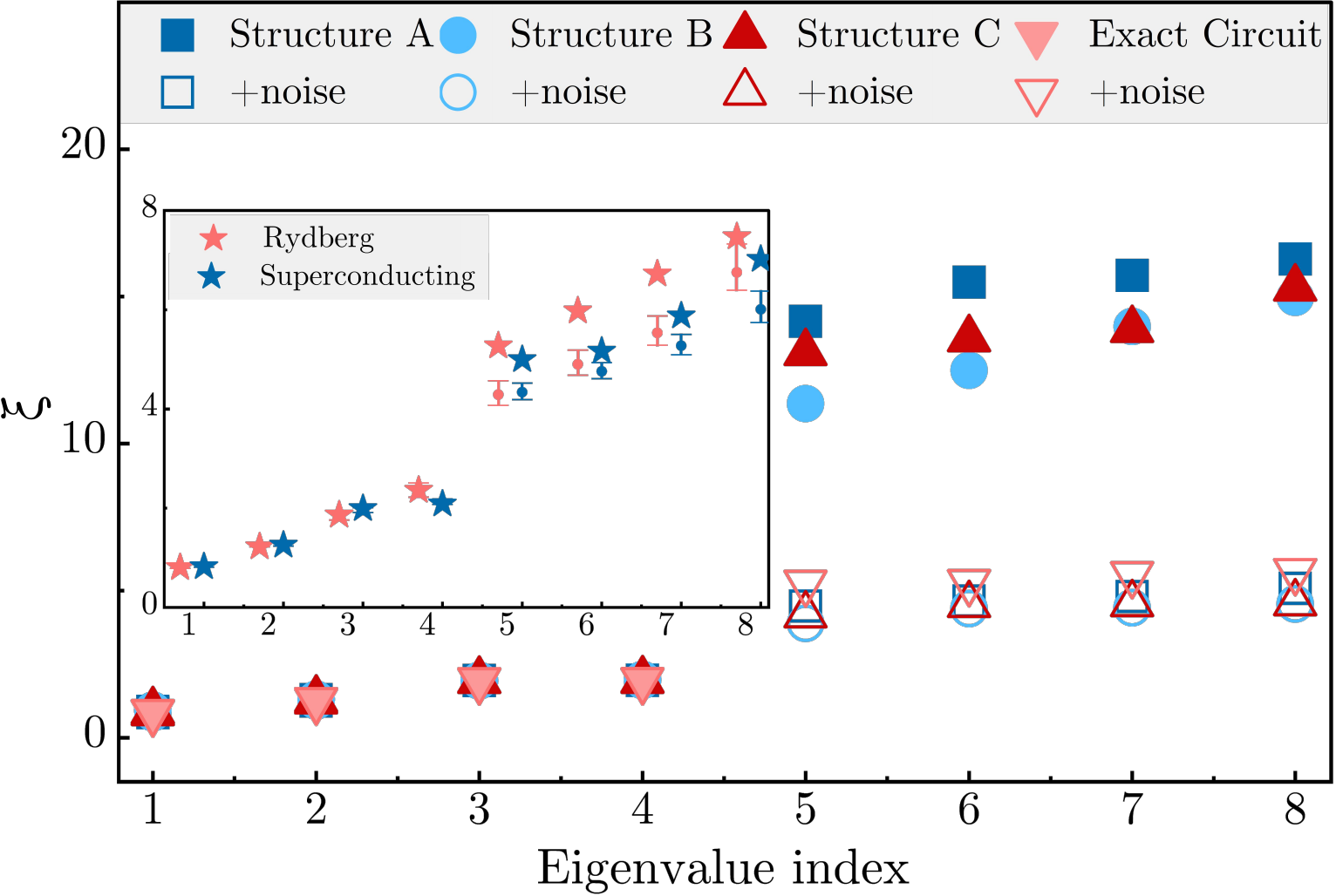}
  \caption{Entanglement spectra for the exact circuit and different VQC structures under noise. Inset: Entanglement spectra obtained with the optimal control method. For the superconducting platform, all results are computed with an evolution time of $0.1\ \mu\mathrm{s}$, while an evolution time of $2\ \mu\mathrm{s}$ is adopted for the Rydberg system. Star symbols in the inset denote noise-free reference results, and all error bars with their corresponding mean values are derived under fixed noise strengths: $20 \%$ for superconducting qubits and $5 \%$ for Rydberg atoms.}
  \label{fig:spec_comparison}
\end{figure}

To characterize the topological properties of the prepared quantum states, we analyze the entanglement spectrum (ES), which provides a more refined probe than scalar quantities such as fidelity or entanglement entropy. The ES is defined via the relation $\xi_i = -\ln(\lambda_i)$, where $\{\lambda_i\}$ are the eigenvalues of the reduced density matrix $\rho_A$ \cite{PhysRevLett.101.010504}. For FQH states, the universal structure of the low-energy ES, separated by a well-defined entanglement gap, serves as a hallmark of topological order and is in one-to-one correspondence with the edge excitation spectrum \cite{PhysRevLett.101.010504,PhysRevB.84.205136}.

For the $7$-qubit system, we partition the first three qubits as subsystem $A$ and compute the ES for both the optimized states and the theoretical target state. The results are summarized in Fig.~\ref{fig:spec_comparison}, with the levels $\xi_i$ sorted in ascending order. In the noise-free case, the exact construction circuit accurately reproduces the target spectrum, maintaining a clear entanglement gap that separates the low-lying topological levels of the Laughlin state from other high levels caused by errors in state preparation. Upon introducing noise, the spectral levels exhibit slight shifts and the levels from the state-preparation error go down. Nevertheless, the lowest-lying levels remain in close agreement with the ideal results, and the entanglement gap remains clearly discernible. This persistence indicates that the essential topological structure of the Laughlin state is preserved despite the presence of experimental noise.

Regarding the VQC approach, all considered circuit structures successfully capture the low-lying part of the target spectrum. Even under the noise levels, the core topological features remain robust, with the level distribution and the entanglement gap being faithfully reconstructed. These findings demonstrate that the variational protocols are capable of preparing states that inherit the characteristic topological signatures of the FQH phase.

The inset of Fig.~\ref{fig:spec_comparison} displays the ES obtained from optimal control schemes, where star symbols denote the noise-free reference results and error bars correspond to mean values under fixed hardware noise strengths, $20\%$ for the superconducting platform and $5\%$ for the Rydberg atom platform. These noise levels are specifically selected as they represent the critical regime where system fidelity begins to degrade noticeably yet remains within an experimentally meaningful and acceptable range, with corresponding state fidelities of $92.01\%$ for the superconducting platform and $89.58\%$ for the Rydberg atom platform. Both platforms reproduce the low-lying structure of the target topological spectrum with high fidelity, showing only minor deviations due to control errors. These results further confirm the robustness of the optimized protocols and their ability to preserve the underlying topological properties across different hardware implementations.

\begin{figure*}
    \centering
    \includegraphics[width=\linewidth]{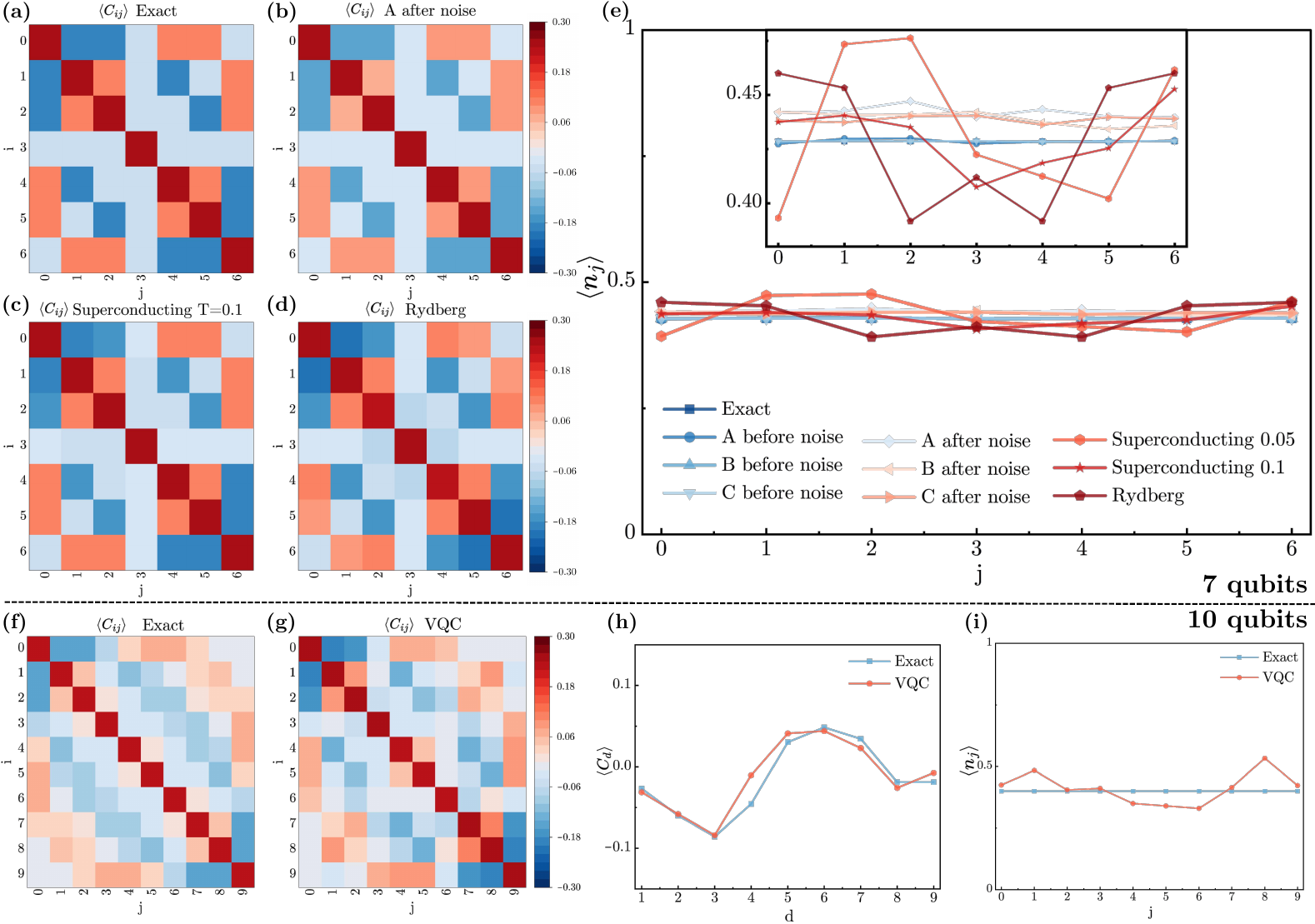}
    \caption{Characterization of prepared states for 7‑qubit and 10‑qubit systems.  
For 7 qubits: (a)–(d) two‑point correlation function \(\langle C_{ij} \rangle\) of the exact target state and states prepared via VQC (with gate noise) and optimal control (superconducting and Rydberg atom schemes); (e) orbital‑resolved average particle number density \(\langle n_j \rangle\) for VQC and optimal control schemes (inset: magnified fine density variation).  
For 10 qubits: (f) correlation matrix from exact diagonalization, (g) VQC simulation; (h) distance‑dependent correlation function; (i) orbital‑dependent occupation numbers.}
    \label{fig:entanglement}
\end{figure*}

\subsection{Density structure and many-body correlation characterization}
To further validate the successful preparation of the Laughlin state beyond fidelity and topological entanglement signatures, we characterize the prepared states via two experimentally accessible observables: the orbital-resolved particle number density and the two-point correlation function. While these quantities do not serve as an order parameter for the Laughlin topological order, they carry fingerprints of the many-body correlations of the Laughlin state, and can be readily measured on near-term quantum hardware via computational basis sampling and minimal quantum state tomography, avoiding the large measurement overhead required for full state tomography or entanglement spectrum reconstruction.

We first compute the orbital occupation number $\langle n_j \rangle$, which maps the real-space density profile of the many-body state [Fig.~\ref{fig:entanglement}(e)]. As an incompressible quantum liquid, the Laughlin state should possess a nearly uniform featureless bulk density. All three noise-free VQC architectures yield density profiles almost identical to the exact result. Under depolarizing gate noise, the VQC results show only minor deviations. Optimal control protocols for both superconducting and Rydberg platforms also accurately reproduce the target density profile, with high agreement with the exact solution maintained even in the presence of control amplitude noise [inset of Fig.~\ref{fig:entanglement}(e)]. These results confirm that all our protocols preserve the density distribution of the FQH state under realistic NISQ noise conditions, demonstrating their robustness and experimental viability.

We further evaluate the two-point correlation function $\langle C_{ij} \rangle = \langle n_i n_j \rangle-\langle n_i \rangle\langle  n_j \rangle$, which encodes the many-body spatial correlations of the FQH liquid. The diagonal elements of the correlation matrix characterize the on-site particle number fluctuations, while the off-diagonal elements quantify the non-local quantum coherence between different orbitals. As shown in Fig.~\ref{fig:entanglement}(a)-(d), the exact target state [Fig.~\ref{fig:entanglement}(a)] exhibits the characteristic correlation pattern of the $\nu=1/3$ Laughlin state: strong negative short-range correlations and rapid decay of long-range correlations, a signature of an incompressible homogeneous quantum liquid with repulsive inter-particle interactions. The noisy VQC result [Fig.~\ref{fig:entanglement}(b)], superconducting optimal control result [Fig.~\ref{fig:entanglement}(c)], and Rydberg analog result [Fig.~\ref{fig:entanglement}(d)] all show virtually indistinguishable correlation matrices from the exact target, with both the diagonal and off-diagonal elements in excellent quantitative agreement. This confirms that our protocols not only achieve high state fidelity and preserve the topological entanglement spectrum, but also faithfully reproduce the intrinsic many-body correlations of the FQH state, ensuring the prepared states are physically identical to the target topological phase.

For the $10$-qubit \( \nu=1/3 \) Laughlin state, as shown in Fig.~\ref{fig:entanglement}(f)-(i), compared to the exact state, the experimentally prepared state exhibits small fluctuations in occupation numbers, the overall trend of the correlation coefficient matrix is consistent, and the distance-dependent correlation analysis also shows good agreement.

\section{Discussion and Outlook}
In this work, we have investigated the preparation of FQH states on quantum computers from three complementary perspectives: direct circuit construction, variational quantum circuits, and optimal control. By focusing on the $\nu=1/3$ Laughlin state on the sphere, we move beyond the commonly studied thin-torus or quasi-one-dimensional limits and address the preparation of a genuine two-dimensional topological state with nontrivial entanglement structure.

Our results highlight the importance of exploiting the intrinsic structure of the target state. In the direct construction, the constrained squeezing structure of the FQH wavefunction enables a substantial reduction in circuit complexity, particularly in the number of required multi-controlled gates. Compared to generic state-preparation schemes based on binary-tree decompositions, the effective control depth is significantly lowered, making the approach more amenable to near-term devices. The variational approach, on the other hand, provides a flexible and hardware-adaptable framework, but generally requires deeper circuits and a larger number of two-qubit gates to achieve comparable fidelity. Our analysis further demonstrates that the choice of qubit connectivity plays a crucial role in variational performance, emphasizing the need for co-design between ansatz structure and hardware topology. Finally, the optimal control method offers a fundamentally different route by directly engineering the target state through nonadiabatic dynamics, avoiding the limitations of adiabatic protocols and enabling fast state preparation under realistic control constraints.

Despite these advances, several remaining open questions require further studies in future works. First, while we have demonstrated high-fidelity preparation for the seven-qubit system, extending these methods to larger systems increases the complexity of both circuit design and optimization. For instance, in the ten-qubit case, we observe that the number of two-qubit gates rises from $14$ to $29$. More importantly, an additional $16$ multi-qubit gates are needed. Therefore, developing an informed estimate for the scaling law is a key task for future studies.
Second, our results highlight an intrinsic trade-off between circuit depth, qubit overhead, and control complexity. While parallel circuit constructions can achieve reduced depth at the expense of additional qubit resources, our approach prioritizes minimizing qubit requirements and multi-controlled gate complexity, making it more suitable for near-term quantum devices. However, our current constructions are not guaranteed to be optimal in terms of circuit depth, gate count, or control resources.
Developing more efficient preparation strategies that further exploit the structure of FQH states, such as their sparsity, symmetries, and underlying algebraic properties, remains an important direction for future work. Third, the design principles of the circuits presented here can be readily generalized to study FQH states with alternative filling factors, for instance the $\nu=5/2$ Moore-Read state that features non-Abelian anyons.
Finally, an important future direction is the experimental realization of these protocols. In this regard, the exact scheme can be further optimized on the circuit compilation level to further compress the circuit depth. 
Incorporating realistic noise models, gate errors, and decoherence effects will be essential for assessing the feasibility of preparing FQH states on current quantum hardware. In addition, extending these methods to probe dynamical properties, excitations~\cite{Kirmani2022}, and topological invariants of FQH systems may open new avenues for exploring strongly correlated quantum matter on programmable quantum platforms.

\begin{acknowledgments}

This work is supported by the National Key Research and Development 
of China (Grant Nos. 2021YFA1402001) and the National Natural Science 
Foundation of China (NSFC) (Grant Nos. 12375007). Z.~L. was supported by 
the National Key Research and Development Program of China (Grant No. 2021YFA1401902).

\end{acknowledgments}

\bibliography{references}



\clearpage

\appendix

\section{Minimal example of our method}\label{q3example}
\begin{figure}
    \centering
      \includegraphics[width=\linewidth]{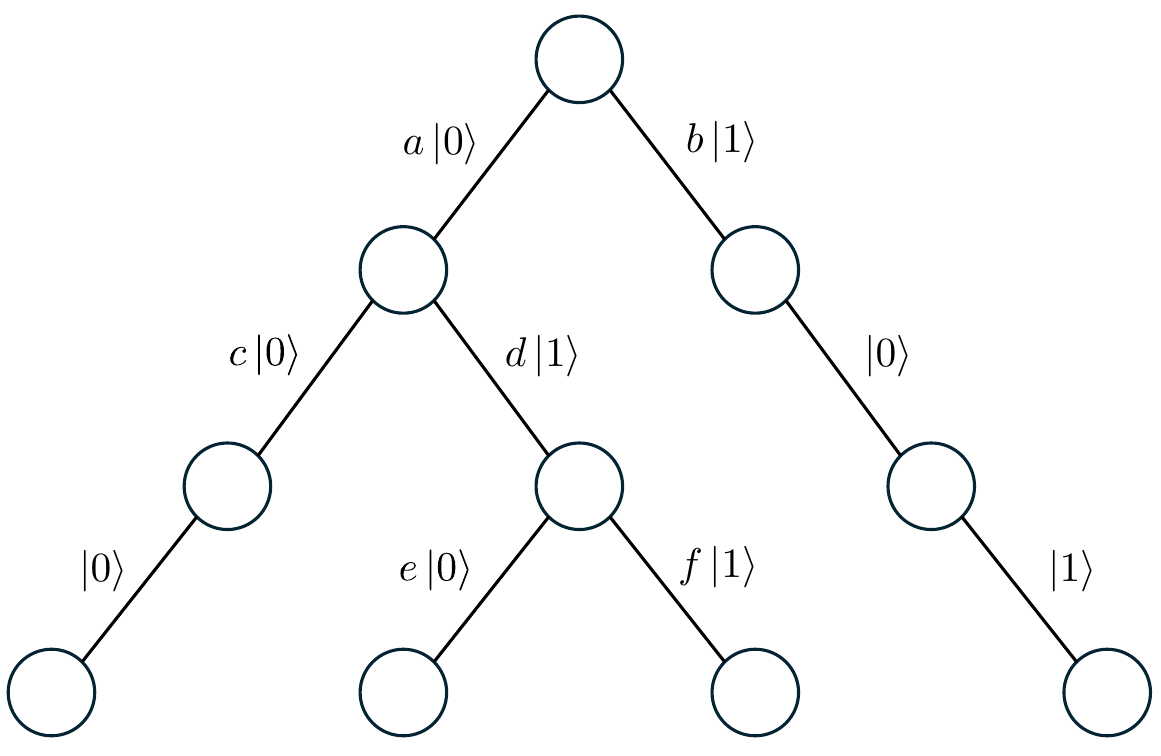}
      \caption{Binary tree representation of the $3$-qubit target state in Eq.~\ref{q3_target_state}. }
      \label{fig:q3_binary_tree}
\end{figure}

In this section, we provide a simple example to illustrate how to directly construct quantum circuits for the $7$-qubit and $10$-qubit FQH states shown in Fig.~\ref{circuit} and Fig.~\ref{fig:circuit10}.
We begin with a three-qubit example. Suppose the target state is given by 
\begin{equation}
    \ket{\psi_t} = \alpha\ket{000}+\beta\ket{010}+\gamma\ket{011}+\delta\ket{101}, \label{q3_target_state}
\end{equation}
where $\alpha,\beta,\gamma,\delta$ are free parameters satisfying the normalization condition 
\begin{equation}
    |\alpha|^2+|\beta|^2+|\gamma|^2+|\delta|^2=1. 
\end{equation}
The state $\ket{\psi_t}$ can be represented by a binary tree with weighted edges, as shown in Fig.~\ref{fig:q3_binary_tree}, where the parameters $a, b, c, d, e, f$ are determined by $\alpha,\beta,\gamma,\delta$ 
via solving the follwoing equations, 
\begin{eqnarray}
    ac &=& \alpha,\\
    ade &=&\beta,\\
    adf &=& \gamma,\\
    b &=& \delta,
\end{eqnarray}
with three normalization constraints, 
\begin{eqnarray}
    |a|^2+|b|^2&=&1,\\
    |c|^2+|d|^2&=&1,\\
    |e|^2+|f|^2&=&1. 
\end{eqnarray}

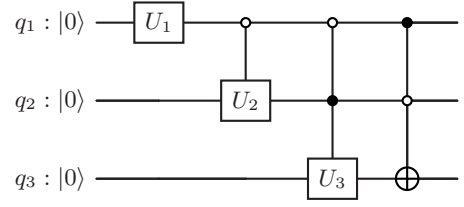
\begin{figure}
    \centering
        \begin{quantikz}
           \lstick{$q_1:\ket{0}$}  & \gate{U_1} & \octrl{1} & \octrl{1}  & \ctrl{2}  & \qw \\
            \lstick{$q_2:\ket{0}$} & \qw       & \gate{U_2} & \ctrl{1}   & \octrl{1} & \qw \\
            \lstick{$q_3:\ket{0}$} & \qw       & \qw        & \gate{U_3} & \targ{}   & \qw
        \end{quantikz}
    \caption{A circuit that prepares the $3$-qubit target state without any simplification. 
    Here $U_1,U_2,U_3$ transforms $\ket{0}$ to $a\ket{0}+b\ket{1}, c\ket{0}+d\ket{1}$ and $e\ket{0}+f\ket{1}$, respectively. }
    \label{q3_circuit}
\end{figure}

In this representation, the state of the $n$-th qubit is conditioned on the states of the preceding $n-1$ qubits.
For example, consider the preparation of the second qubit. Starting from the initial state $\ket{0}$, the first qubit is first transformed into $a\ket{0}+b\ket{1}$. According to the binary tree, when the first qubit is in $\ket{0}$, the second qubit should be transformed as $\ket{0}\to c\ket{0}+d\ket{1}$, while no operation is required when the first qubit is in $\ket{1}$. This conditional transformation can be implemented using a controlled unitary gate, where the first qubit acts as the control and the second qubit as the target.
The procedure for the third qubit follows a similar logic. When the first two qubits are in the state $\ket{01}$, the third qubit should be transformed as $\ket{0}\to e\ket{0}+f\ket{1}$, which requires a two-controlled unitary gate acting on the third qubit. In addition, when the first two qubits are in $\ket{10}$, the third qubit needs to be flipped from $\ket{0}$ to $\ket{1}$. The resulting circuit constructed in this manner is shown in Fig.~\ref{q3_circuit}.
From this construction, we observe that, in general, the operation on the $n$-th qubit depends on all preceding $n-1$ qubits, leading to the requirement of multi-controlled gates of the form $C^{n-1}U$. Such gates quickly become costly and challenging to implement on current quantum hardware.

\begin{figure}
    \centering
        \begin{quantikz}
           \lstick{$q_1:\ket{0}$}  & \gate{U_1} & \octrl{1} & \qw        & \ctrl{2}  & \qw \\
            \lstick{$q_2:\ket{0}$} & \qw       & \gate{U_2} & \ctrl{1}   & \qw & \qw \\
            \lstick{$q_3:\ket{0}$} & \qw       & \qw        & \gate{U_3} & \targ{}   & \qw
        \end{quantikz}
    \caption{A simplified circuit that prepares the $3$-qubit target state. 
    Here $U_1,U_2,U_3$ transforms $\ket{0}$ to $a\ket{0}+b\ket{1}, c\ket{0}+d\ket{1}$ and $e\ket{0}+f\ket{1}$, respectively. }
    \label{q3_circuit_simplified}
\end{figure}
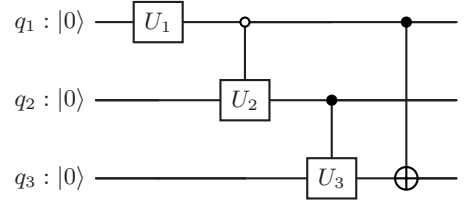

However, the specific structure of $\ket{\psi_t}$ allows for a further simplification of the circuit. Notably, the transformation on the third qubit does not depend on the full configuration of the first two qubits. For instance, the operation associated with the branch $\ket{01}$ can be implemented by conditioning only on the second qubit being in $\ket{1}$, without requiring explicit control on the first qubit. As a result, the corresponding $C^2U$ gate can be reduced to a standard controlled-$U$ gate. Similarly, the bit-flip operation required for the branch $\ket{10}$ does not depend on both qubits simultaneously, but only on the first qubit being in $\ket{1}$. Therefore, this operation can also be implemented using a single control qubit. With these simplifications, the circuit no longer requires costly two-controlled gates, and can be implemented entirely using single-controlled operations, as shown in Fig.~\ref{q3_circuit_simplified}. This example illustrates how the intrinsic structure of the target state can be exploited to significantly reduce the complexity of the circuit.

\section{Digital circuit and VQC results for $10$-qubit $\nu=1/3$ FQH state}\label{fqh10_circuit}
\begin{table}[]
\begin{tabular}{@{}ccccccccllll@{}}
\toprule
bitstring & $q_0$ & $q_1$ & $q_2$ & $q_3$ & $q_4$ & $q_5$ & $q_6$ & $q_7$ & $q_8$ & $q_9$ & amplitude                 \\ \midrule
1         & $1$   & $0$   & $0$   & $1$   & $0$   & $0$   & $1$   & $0$   & $0$   & $1$   & $-\frac{2}{5}\sqrt{\frac{3}{17}}$  \\
2         & $1$   & $0$   & $0$   & $1$   & $0$   & $0$   & $0$   & $1$   & $1$   & $0$   & $\frac{2}{5}\sqrt{\frac{7}{17}}$  \\
3         & $1$   & $0$   & $0$   & $0$   & $1$   & $1$   & $0$   & $0$   & $0$   & $1$   & $\frac{4}{5}\sqrt{\frac{3}{17}}$  \\
4         & $1$   & $0$   & $0$   & $0$   & $1$   & $0$   & $1$   & $0$   & $1$   & $0$   & $-\frac{4}{5}\sqrt{\frac{2}{17}}$  \\
5         & $1$   & $0$   & $0$   & $0$   & $0$   & $1$   & $1$   & $1$   & $0$   & $0$   & $\sqrt{\frac{2}{17}}$             \\ 
6         & $0$   & $1$   & $1$   & $0$   & $0$   & $0$   & $1$   & $0$   & $0$   & $1$   & $\frac{2}{5}\sqrt{\frac{7}{17}}$  \\ 
7         & $0$   & $1$   & $1$   & $0$   & $0$   & $0$   & $0$   & $1$   & $1$   & $0$   & $-\frac{14}{15}\sqrt{\frac{3}{17}}$\\ 
8         & $0$   & $1$   & $0$   & $1$   & $0$   & $1$   & $0$   & $0$   & $0$   & $1$   & $-\frac{4}{5}\sqrt{\frac{2}{17}}$  \\ 
9         & $0$   & $1$   & $0$   & $1$   & $0$   & $0$   & $1$   & $0$   & $1$   & $0$   & $\frac{8}{15}\sqrt{\frac{3}{17}}$ \\ 
10        & $0$   & $1$   & $0$   & $0$   & $1$   & $1$   & $0$   & $0$   & $1$   & $0$   & $\frac{4}{15}\sqrt{\frac{3}{17}}$ \\ 
11        & $0$   & $1$   & $0$   & $0$   & $1$   & $0$   & $1$   & $1$   & $0$   & $0$   & $-\frac{3}{5}\sqrt{\frac{2}{17}}$  \\ 
12        & $0$   & $0$   & $1$   & $1$   & $1$   & $0$   & $0$   & $0$   & $0$   & $1$   & $\sqrt{\frac{2}{17}}$             \\ 
13        & $0$   & $0$   & $1$   & $1$   & $0$   & $1$   & $0$   & $0$   & $1$   & $0$   & $-\frac{3}{5}\sqrt{\frac{2}{17}}$  \\ 
14        & $0$   & $0$   & $1$   & $1$   & $0$   & $0$   & $1$   & $1$   & $0$   & $0$   & $\frac{1}{15}\sqrt{\frac{3}{17}}$  \\ 
15        & $0$   & $0$   & $1$   & $0$   & $1$   & $1$   & $0$   & $1$   & $0$   & $0$   & $\frac{1}{3}\sqrt{\frac{3}{17}}$  \\ 
16        & $0$   & $0$   & $0$   & $1$   & $1$   & $1$   & $1$   & $0$   & $0$   & $0$   & $-\frac{1}{3}\sqrt{\frac{3}{17}}$  \\ 
\bottomrule
\end{tabular}
\caption{Configuration of 10-qubit FQH state we prepare in Fig.~\ref{fig:circuit10}.  \label{FQH10}}
\end{table}

\begin{figure*}
    \includegraphics[width=0.95\linewidth]{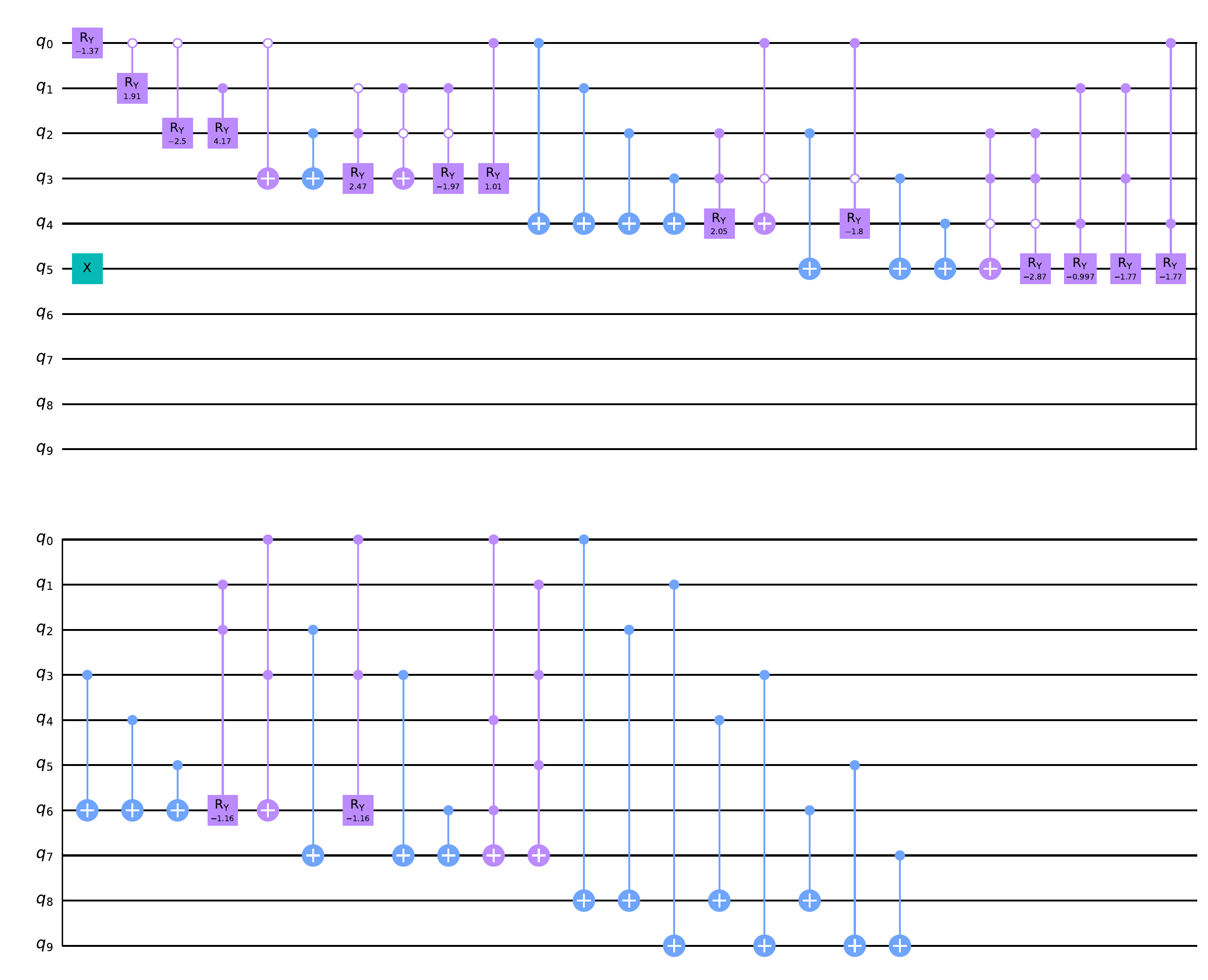}
    \caption{Circuit that realizes target state defined in Table.~\ref{FQH10}, depicted using qiskit~\cite{qiskit2024}. 
    The circuit is initialized in $\ket{0}^{\otimes 10}$.}
    \label{fig:circuit10}
\end{figure*}

The bitstrings and the correspoding amplitudes of $10$-qubit $\nu=1/3$ laughlin states are listed in Table.~\ref{FQH10}. 
The circuit is constructed as discussed in Sec.~III and shown in Fig.~\ref{fig:circuit10}. 

\begin{figure}[h]
    \includegraphics[width=\linewidth]{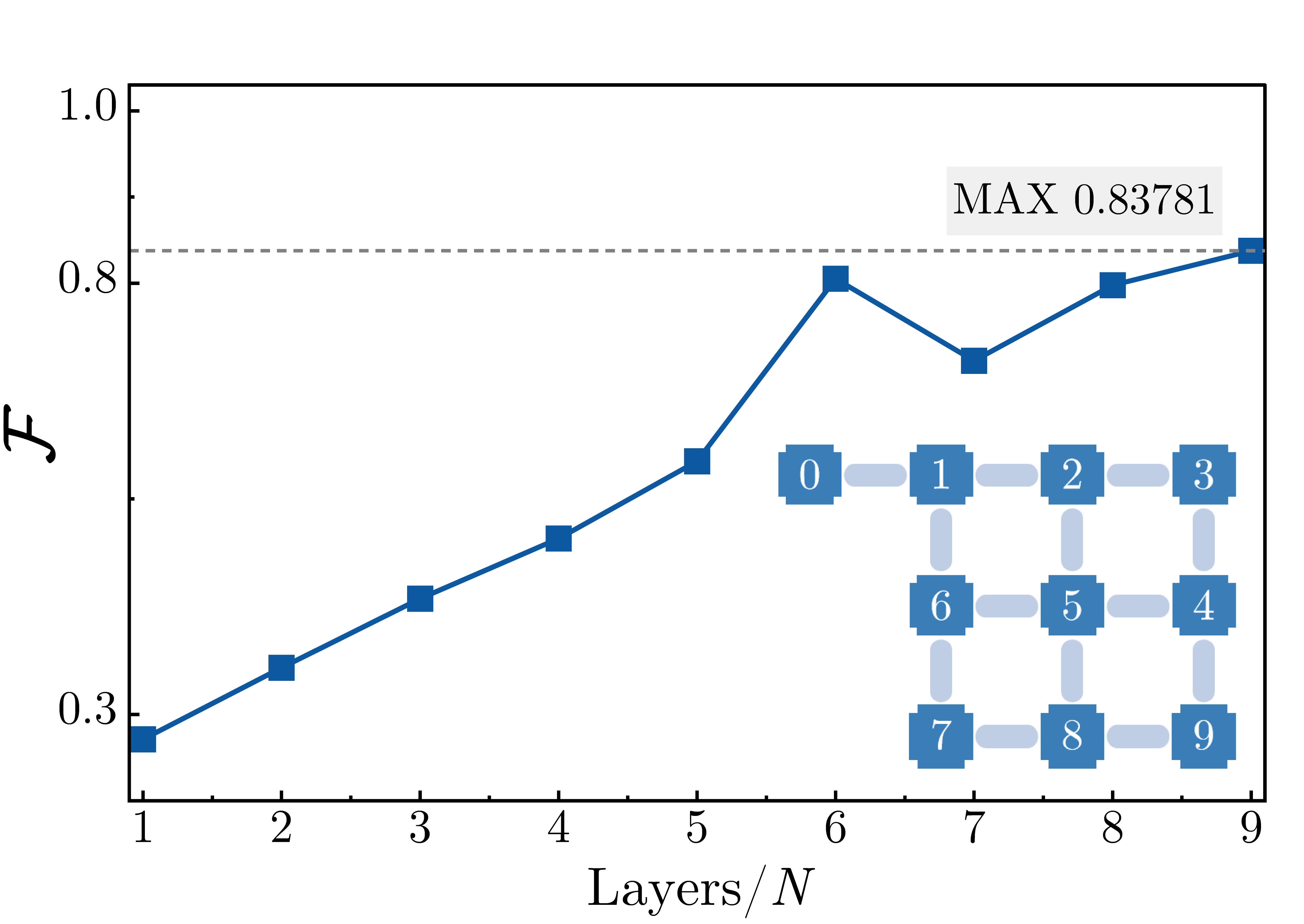}
    \caption{VQC results for $10$-qubit $\nu=1/3$ laughlin states.}
    \label{fig:VQC-for-10}
\end{figure}

For the 10-qubit FQH state, the exact circuit construction requires approximately 150 CNOT gates, a dramatic increase compared to only 14 CNOT gates for the 7-qubit case. This substantial growth in circuit complexity and two-qubit gate count means that the VQC method also requires a corresponding increase in the number of ansatz layers to maintain high state-preparation fidelity.

For variational state preparation of the $10$-qubit system, the increased Hilbert space dimension and circuit complexity lead to a notably lower achievable fidelity compared to the 7-qubit case. Nevertheless, by increasing the number of ansatz layers, we achieve a state preparation fidelity of 83.78\% via the VQC approach, with the fidelity convergence curve shown in Fig.~\ref{fig:VQC-for-10}. The topological properties, density profile, and many-body correlations of the prepared $10$-qubit state are analyzed and discussed in detail in Sec.~IV.

\begin{figure*}
    \centering
    \includegraphics[width=0.95\linewidth]{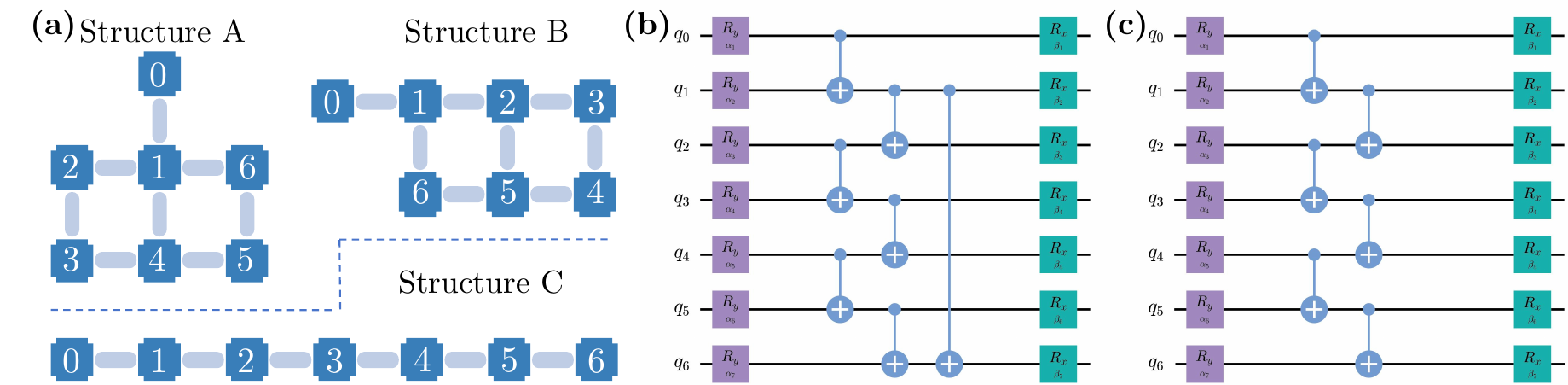}
    \caption{Topological structures and ciruits of the three VQC configurations.(a) Topological structures of Structure B and Structure C. (b) One-layer VQC in Structure B. (c) One-layer VQC in Structure C.}
    \label{fig:B-C}
\end{figure*}

\section{VQC Optimization Details}
\label{app:vqe_parameters}
Three different configurations of the VQC are constructed, each corresponding to a distinct connectivity pattern of CNOT gates within a single layer. Structure A is described in detail in the main text. Other configurations are described in this appendix.Their topological structures are illustrated in Fig.~\ref{fig:B-C}(a).

The qubit orderings of the three structures are all computed based on the degree of correlation between different singlet states, and it has also been verified through numerical simulations that this configuration most readily yields high-fidelity results.

For Structure A, CNOT gates are formed between adjacent qubits, with phase and bit-flip gates of a fixed rotation angle applied to all qubits in each single layer; this layout additionally makes qubit 1 adjacent to qubit 4 and qubit 6 to accommodate extra CNOT gates. 
\begin{table}[h]
\centering
\caption{Optimized rotation parameters for VQC with 4 composite layers}
\begin{tabular}{ccc | ccc}
\hline
Layer & $R_y$ ($\boldsymbol{\alpha}$) & $R_x$ ($\boldsymbol{\beta}$) & Layer & $R_y$ ($\boldsymbol{\alpha}$) & $R_x$ ($\boldsymbol{\beta}$) \\
\hline
 & $q_0$: $2.0000\pi$ & $2.5000\pi$ & & $q_0$: $2.5000\pi$ & $2.5000\pi$ \\
 & $q_1$: $2.8273\pi$ & $0.6639\pi$ & & $q_1$: $2.0000\pi$ & $3.0000\pi$  \\
 & $q_2$: $3.5000\pi$ & $1.9947\pi$ & & $q_2$: $0.5000\pi$ & $1.0000\pi$ \\
\textbf{1} & $q_3$: $2.5000\pi$ & $2.2277\pi$ & \textbf{3} & $q_3$: $2.0000\pi$ & $2.5000\pi$ \\
 & $q_4$: $1.4999\pi$ & $3.2474\pi$ & & $q_4$: $2.5000\pi$ & $3.5000\pi$ \\
 & $q_5$: $3.2677\pi$ & $0.2355\pi$ & & $q_5$: $0.0003\pi$ & $1.0000\pi$ \\
 & $q_6$: $0.0000\pi$ & $2.7180\pi$ & & $q_6$: $0.5000\pi$ & $1.0000\pi$ \\
\hline
 & $q_0$: $2.6207\pi$ & $3.9991\pi$ & & $q_0$: $2.0009\pi$ & $1.0000\pi$ \\
 & $q_1$: $3.5000\pi$ & $0.0000\pi$ & & $q_1$: $0.0000\pi$ & $0.0000\pi$ \\
 & $q_2$: $2.0000\pi$ & $0.5249\pi$ & & $q_2$: $0.0000\pi$ & $3.0000\pi$ \\
\textbf{2} & $q_3$: $1.0000\pi$ & $3.1959\pi$ & \textbf{4} & $q_3$: $2.5000\pi$ & $0.5000\pi$ \\
 & $q_4$: $0.0000\pi$ & $3.0000\pi$ & & $q_4$: $2.0000\pi$ & $1.5000\pi$ \\
 & $q_5$: $2.5000\pi$ & $3.0000\pi$ & & $q_5$: $1.0000\pi$ & $0.5000\pi$ \\
 & $q_6$: $2.5000\pi$ & $1.1959\pi$ & & $q_6$: $1.5000\pi$ & $0.5000\pi$ \\
\hline
\end{tabular}
\label{tab:vqe_parameters}
\end{table}

The single-layer schematics of Structure B and Structure C are presented in Fig.~\ref{fig:B-C}(b)(c). For comparison, Structure B only creates adjacency between qubit 1 and qubit 6, while Structure C cannot provide any additional CNOT gates.

Among the optimized data, a comparison of fidelity can be seen in the main text (Fig.~\ref{fig:topology}). The other configurations achieve lower fidelity than Configuration A because they have fewer forms of inter-qubit control, which necessitates deeper circuit layers to meet the entanglement requirements of the target state, resulting in relatively lower fidelity.

Here we list the optimization coefficient results for Structure A at 4 layers in Table.~\ref{tab:vqe_parameters}.

\end{document}